\documentstyle[12pt]{article}
\begin{document}
\title{
\rightline {\small \bf MADPH-95-872, UG-FT-51/95}
\vskip 1cm
CP violation in the lepton sector with Majorana neutrinos}
\author{$\mbox{\rm F. del Aguila}^{1}$ and $\mbox{\rm M. Zra{\l }ek}^{2}$\\
\small $^1$ Depto. de F\'\i sica Te\'orica y del Cosmos,\\
\small Universidad de Granada, E-18071 Granada, Spain \\
\small $^2$ Field Theory
and Particle Physics Dept.\\
\small University of Silesia, 40-007 Katowice, Poland}
\maketitle
\thispagestyle{empty}
\begin{abstract}
We study CP violation in the lepton sector in extended models with
right-handed neutrinos, without and with left-right symmetry, and
with arbitrary mass terms. We find the
conditions which must be satisfied by the neutrino and charged lepton
mass matrices for CP conservation. These constraints, which are independent
of the choice of weak basis, are proven to be also sufficient in
simple cases. This invariant formulation makes apparent
the necessary requirements for CP violation, as well as the size
of CP violating effects. As an example, we show that CP violation
can be much larger in left-right symmetric models than in models
with only additional right-handed neutrinos, {\it i.e.}, without
right-handed currents.
\end{abstract}
\newpage
{\bf 1. Introduction}\\

The origin of CP violation is still an open problem in particle physics.
In the standard model CP violation is related to the mixing between flavour
and mass eigenstates (Cabbibo-Kobayashi-Maskawa (CKM) mechanism) [1].
In this case three standard families of quarks with non-degenerate masses
must exist. This is known to happen, and small CP violating effects has
been observed in the $K^0-\bar K^0$ system.

Why these effects must be small is also understood. Physical quantities are
independent of the choice of (weak) quark basis. Hence, only weak basis
invariants enter in measurable quantities like cross sections or decay
widths. Two sets of CP symmetry breaking invariants have been constructed
[2,3]. For three families there is only one independent invariant, which
can be chosen to be the determinant of the commutator of $M_uM^{\dagger }_u$
and $M_dM^{\dagger }_d$ [2],
\begin{equation}
\begin{array}{c}
Det [ M_uM^{\dagger }_u, M_dM^{\dagger }_d ] =
-2i\ (m_t^2-m_c^2)(m_t^2-m_u^2)(m_c^2-m_u^2) \\
(m_b^2-m_s^2)(m_b^2-m_d^2)(m_s^2-m_d^2)
\ Im(V_{ud}V_{cs}V^*_{us}V^*_{cd}) ,
\end{array}
\end{equation}
where $M_{u(d)}$ is the up (down) quark mass matrix in a weak current
eigenstate basis, $m_i$ is the mass of the quark $i$, and $V_{ij}$ is
the $ij$ entry of the CKM matrix; or the trace of the triple product
of the commutator [3],
\begin{equation}
Tr [ M_uM^{\dagger }_u, M_dM^{\dagger }_d ] ^3 =
3\ Det [ M_uM^{\dagger }_u, M_dM^{\dagger }_d ] .
\end{equation}
As a consequence, in the standard model all CP violating effects are
proportional to
\begin{equation}
\delta _{KM} =
Im(V_{ud}V_{cs}V^*_{us}V^*_{cd}) .
\end{equation}
Using the unitarity of the CKM matrix we can write
\begin{equation}
| \delta _{KM} | =
| Im(V_{ub}V_{cs}V^*_{us}V^*_{cb}) | ,
\end{equation}
and substituting the experimental values of $|V_{ij}|$ in Eq. (4) it can
be shown that
\begin{equation}
| \delta _{KM} | \leq
10^{-4} ;
\end{equation}
and then that all CP violating effects are small [4]. This is usually
summarized saying that CP violation has been only observed in the
$K^0-\bar K^0$ system and that it is small due to the small mixing
angles $|V_{ij}|$ in Eq. (4).

No CP violation has been observed in the lepton sector. This may be
related to the smallness (or vanishing) of the masses of the electron,
muon and tau neutrinos. LEP results exclude the existence of more
than three left-handed neutrinos with masses practically up to the
$Z^0$ mass. Neutrinos with larger masses can exist and are predicted
in many models beyond the standard model. How large can CP violation
be in the lepton sector with heavy (Majorana) neutrinos has not been
fully analysed to our knowledge. Only the number of CP violating phases
for different neutrino contents has been calculated [5].

As for quarks, the construction of a set of
CP violating invariants for lepton mass matrices shall allow for:
\begin{itemize}
\item deciding more easily in any weak lepton basis if CP is conserved;
\item understanding the origin of CP violation if for any (physical)
reason a definite (class of) model(s) is distinguished;
\item motivating model building; and
\item obtaining the size of CP violating effects, because
all physical quantities are proportional to weak basis
invariants and then, knowing a set of neccesary and sufficient
invariant constraints for CP conservation stands for knowing
the possible factors suppressing the CP violating observables.
\end{itemize}

In this paper we discuss extended models with an arbitrary number of
right-handed neutrinos and standard lepton families, without (Section 2)
and with (Section 3) left-right
symmetry,. We find necessary conditions for CP conservation for
arbitrary lepton mass matrices. These, which are independent of the choice
of weak basis, are proven to be sufficient in simple cases. Using
these invariant conditions we address the question of how large can
CP violation effects be for models with and without left-right
symmetry (Section 4). (We do not discuss other possible CP violating effects
mediated by Higgses.) \\

{\bf 2. Extended models with right-handed neutrinos}\\

We proceed analogously to the quark case [3]: we state the conditions
for CP conservation in the lepton sector satisfied in any weak basis,
we count the number of CP breaking phases in a particular basis [5],
and we construct invariant conditions for CP conservation, proving in
some simple cases that these are necessary and sufficient, what
constitutes our main result. \\

{\bf 2.1. CP invariance}\\

The gauge interactions in extended electroweak models with $n_R$
right-handed neutrinos (and $n_L$ standard lepton families)
are written in any weak basis as in the standard model [4]:
\begin{equation}
{\cal L}_{CC} = - \frac{g}{\sqrt{2}} \bar {\nu}_L
\gamma^{\mu} l_L W^+_{\mu} + h.c.,
\end{equation}
and
\begin{equation}
\begin{array}{c}
{\cal L}_{NC} = e \ ( \bar l_L \gamma^{\mu} l_L +
\bar l_R \gamma^{\mu} l_R ) \ A_{\mu} \\
+ \ \frac{g}{2 cos \theta _W} \ \left( - \bar {\nu}_L \gamma^{\mu} \nu _L
+ ( 1 - 2 sin^2 \theta _W ) \ \bar l_L \gamma^{\mu} l_L
- 2 sin^2 \theta _W \ \bar l_R \gamma^{\mu} l_R \right) \ Z_{\mu},
\end{array}
\end{equation}
where $l_{L,R}$ are two $n_L$ component vectors in flavour space
describing the $n_L$ left-handed and $n_L$ right-handed charged
leptons, and $\nu _L$ is a $n_L$ component vector describing the
$n_L$ left-handed neutrinos. ${\cal L}_{CC,NC}$ are left invariant
by any CP transformation
\begin{equation}
\begin{array}{c}
\nu_{L} \longrightarrow V_{L}C{\nu^{*}_{L}}, \\
l_L \longrightarrow V_{L}C{l^{*}_{L}}, \\
l_{R} \longrightarrow {V^{l}_{R}}C{l^{*}_{R}},
\end{array}
\end{equation}
where $V_L$ and $V_R^l$ are $n_L \times n_L$ unitary matrices,
and $C$ is the Dirac charge conjugation matrix. Similarly a general
CP transformation on the right-handed neutrinos reads
\begin{equation}
\nu_{R} \longrightarrow {V^{\nu}_{R}}C{\nu^{*}_{R}},
\end{equation}
where $V_R^{\nu}$ is a $n_R \times n_R$ unitary matrix. Then CP is
conserved if there exist transformations (8) and (9) that left invariant
the mass terms [6]
\begin{equation}
\begin{array}{c}
{\cal L}_{mass} = - (\bar l_L M_l l_R + \bar l_R M^{\dagger}_l l_L) \\
- \frac{1}{2} (\bar {\psi}_L M_{\nu} \psi _R + \bar {\psi}_R M^*_{\nu}
\psi _L),
\end{array}
\end{equation}
where $M_l$ is a $n_L \times n_L$ complex matrix,
$\psi _{L,R}$ are $n_L+n_R$ component vectors describing the $n_L$ left-handed
and $n_R$ right-handed neutrinos,
\begin{displaymath}
\psi _L = { \nu _L \choose \nu _L^c = i \gamma ^2 \nu ^*_R }, \
\psi _R = { \nu _R^c = i\gamma ^2 \nu ^*_L \choose \nu _R },
\end{displaymath}
and $M_{\nu}$ is a $(n_L+n_R) \times (n_L+n_R)$ complex
symmetric matrix,
\begin{equation}
M_{\nu} =
{\begin{array}{c}
n_L \{ \\ n_R \{
\end{array}}
{ \overbrace{M_L}^{n_L} \ \overbrace{M_D}^{n_R} \choose M^T_D \ M_R }.
\end{equation}
The invariance of (10) under the transformations (8) and (9) implies
\begin{equation}
\begin{array}{c}
{V^{\dagger}_{L}}M_{l}{V^{l}_{R}} = {M^{*}_{l}}, \\
{V^{\dagger}_{L}}M_{L}{V^{*}_{L}} = {M^{*}_{L}}, \\
{V^{\dagger}_{L}}M_{D}{V^{\nu}_{R}} = {M^{*}_{D}}, \\
{V^{\nu \ T}_{R}}M_{R} V^{\nu }_R = {M^{*}_{R}}.
\end{array}
\end{equation}
Thus, CP is conserved in an extended electroweak model with Dirac and
Majorana mass matrices $M_{l,D}, M_{L,R}$ if there exist (unitary) matrices
$V_L, V_R^{l,\nu}$ satisfying (12). The converse is also true, if CP is
conserved, such (unitary) matrices do exist.

We assume that the Yukawa couplings can be related to the mass matrices
and that they satisfy analogous equations. This is often the case for
minimal models. We concentrate on these models, then neglecting possible CP
violating effects mediated by Higgs bosons. \\

{\bf 2.2. Weak basis independence}\\

Eqs. (12) are weak basis independent. In any other weak basis
\begin{equation}
\begin{array}{c}
{\nu '}_{L} = W_{L} \nu_{L}, \\
{l'}_{L} = W_{L}l_{L}, \\
{l'}_{R} = {W^{l}_{R}}l_{R}, \\
{\nu '}_{R} = W^{\nu}_R \nu_{R},
\end{array}
\end{equation}
the mass matrices $M_l, M_{\nu}$ can be written
\begin{equation}
\begin{array}{c}
{M'}_l = W_{L}M_{l}{W^{l\ \dagger}_{R}}, \\
{M'}_L = W_{L}M_{L}{W^T_{L}}, \\
{M'}_D = W_{L}M_{D}{W^{\nu \ \dagger}_{R}}, \\
{M'}_R = {W^{\nu \ *}_{R}}M_{R}{W^{\nu \ \dagger}_{R}},
\end{array}
\end{equation}
and satisfy Eqs. (12) but with unitary matrices
\begin{equation}
\begin{array}{c}
{V'}_L = W_L V_L W^T_L, \\
{V'}^{\ l}_R = W^l_R V^l_R W^{l\ T}_R, \\
{V'}^{\ \nu}_R = W^{\nu}_R V^{\nu}_{R} W^{\nu \ T}_R.
\end{array}
\end{equation}
Although Eqs. (12) are the necessary and sufficient conditions for
CP conservation, they are of little practical use. However, they suggest,
as we show below, how to construct CP invariant constraints which
do not depend explicitly on the unitary matrices $V_L, V_R^{l,\nu}$, and
are then more useful.

First, we introduce a convenient basis to parametrize
CP violation. In this basis we will prove in simple cases afterwards that
the more useful CP invariant constraints are not only necessary but
sufficient. \\

{\bf 2.3. CP conserving gauge interactions in the mass eigenstate basis
and counting of CP breaking phases in the lepton sector}\\

Using the freedom to choose the weak basis we can assume $M_l$ and
$M_R$ diagonal with real positive elements (see Eqs. (14)). In this
basis Eqs. (12) for $M_l$ and $M_R$ imply (for non-degenerate charged
lepton masses and non-degenerate diagonal $M_R$ elements)
\begin{equation}
\begin{array}{c}
(V_L) _{ij} = (V^l_R) _{ij} = e^{{i{\delta_i}}}\delta_{ij}, \\
(V^{\nu}_R) _{ij} = e^{{i{\alpha}_i}}{\delta}_{ij},
\end{array}
\end{equation}
where $\alpha _i$ is equal to $0 \ {\rm or} \ \pi$ (arbitrary) for
non-zero (vanishing)  $(M_R) _{ii}$.
Then Eqs. (12) for $M_L$ and $M_D$ are satisfied and CP conserved if
and only if
\begin{equation}
\begin{array}{c}
(M_L) _{ij} = \  (M^r_L) _{ij} \  e^{ i \frac{1}{2} ( \delta_i +
\delta_j)  } , \\
(M_D) _{ij} = \  (M^r_D) _{ij} \  e^{i\frac{1}{2}(\delta_i - \alpha_j)},
\end{array}
\end{equation}
with $(M^r_L) _{ij}$ and $(M^r_D) _{ij}$ real.
Any extra phase violates CP.
In order to count the CP breaking phases,
let us see how look like the CP conserving gauge
interactions in the mass eigenstate basis.
In the weak basis above,
if CP is conserved (see Eqs. (17)),
\begin{equation}
M_{\nu}  =  {\Sigma \ \ 0 \choose 0 \ \ \Pi}
{M^r_L \ \ \ M^r_D \choose M^{r\ T}_D \
M^r_R} {\Sigma \ \ 0 \choose 0 \ \ \Pi},
\end{equation}
with
\begin{equation}
\begin{array}{c}
\Sigma _{ij}  =  e^{i\frac{\delta_{i}}{2}}\delta_{ij}, \\
\Pi _{ij}  =  e^{-i\frac{\alpha_{i}}{2}}\delta_{ij},
\end{array}
\end{equation}
and $M^r_{L,D}$ real and $M^r_R = \Pi ^2 M_R$, where $M_R$ is a diagonal
matrix with real positive elements. In the mass eigenstate basis with $m_i$
the (positive) neutrino masses,
\begin{equation}
\left( \begin{array}{cccc}
m_1 & & &  \\
& m_2 & &  \\
& & \ddots & \\
& & & m_{n_L+n_R}
\end{array} \right)
=  U^T M_{\nu} U,
\end{equation}
where $U$ is the unitary matrix diagonalizing $M_{\nu}$. For $M_{\nu}$
in Eqs. (18,19) $U$ can be written
\begin{equation}
U = {U^*_L \choose U_R} = {\Sigma ^* \ \ 0 \choose \ 0 \ \ \Pi ^*}
O \Lambda ,
\end{equation}
where $O = {O_L \choose O_R}$ is the real orthogonal
$(n_L+n_R) \times (n_L+n_R)$ matrix diagonalizing
\begin{equation}
{M^r_L \ \ \ M^r_D \choose M^{r\ T}_D \  M^r_R} = O
\left( \begin{array}{cccc}
\epsilon _1 m_1 & & &  \\
& \epsilon _2 m_2 & &  \\
& & \ddots & \\
& & & \epsilon _{n_L+n_R} m_{n_L+n_R}
\end{array} \right)
O^T ,
\end{equation}
with $\epsilon _i = 1 \ {\rm or}\ -1$, and $\Lambda $ the diagonal matrix
guaranteeing positive neutrino masses:
\begin{equation}
\Lambda _{ij} = \delta_{ij} e^{i\frac{\pi}{4} (1-\epsilon _i)}.
\end{equation}
Then, Eq. (6) reads in the mass eigenstate basis (calling the
neutrino mass eigenstates $N_L = U^T \psi _L$)
\begin{equation}
\begin{array}{c}
{\cal L}_{CC} =
- \frac{g}{\sqrt{2}} \bar N_L U^{\dagger}_L
\gamma^{\mu} l_L W^+_{\mu} + h.c. \\
 = - \frac{g}{\sqrt{2}} \bar N_L \Lambda  O^T_L
\gamma^{\mu} \Sigma ^* l_L W^+_{\mu} + h.c. ,
\end{array}
\end{equation}
where $U^*_L = \Sigma ^* O_L \Lambda $ and $O_L$ above are rectangular
matrices corresponding to the first $n_L$ rows of $U$ and $O$,
respectively. The $\Sigma ^*$ phases can be absorbed in the charged
lepton mass eigenstates defining them equal to
$e^{-i\frac{\delta_i}{2}}l_{L i}$. Thus these phases do not violate
CP because they are unphysical. The mixing of the charged current
reduces then to the real matrix $O_L^T$. (The $\Lambda $ phases
can be associated to the neutrino eigenstate definition as we
argue below.) The neutral current lagrangian in Eq. (7) remains
unchanged for the charged lepton mass eigenstates; whereas for the
neutrinos reads
\begin{equation}
\begin{array}{c}
{\cal L}_{NC} = - \frac{g}{2 cos \theta _W} \ \bar N_L U_L^{\dagger}
\gamma^{\mu} U_L N_L \ Z_{\mu} \\
 = - \frac{g}{2 cos \theta _W} \ \bar N_L \Lambda  O_L^T O_L
\gamma^{\mu} \Lambda ^* N_L \ Z_{\mu}.
\end{array}
\end{equation}
In conclusion, the only possible complex factors in the gauge
interactions are the elements of the diagonal matrix $\Lambda $:
\begin{equation}
\Lambda _{kk} = 1(i) \ {\rm for} \ \epsilon _k = 1(-1).
\end{equation}
At any rate CP is conserved if we define the CP parity of the
Majorana neutrino mass eigenstate $k$
\begin{equation}
\eta _{CP} (k) = \epsilon _k i .
\end{equation}
In this case all Majorana neutrinos satisfy
\begin{equation}
N^C \equiv C\bar{N}^T = N .
\end{equation}
Other authors prefer to get rid of the $\Lambda $ phases in the
lagrangian, defining the Majorana neutrino mass eigenstates equal
to $\Lambda ^* N_L $ [7], and considering $\Lambda ^*$ as creation phase
factors [8]. In this case the right-hand side of Eq. (28) is
equal to $-N$ for Majorana neutrinos with $\eta _{CP} = - i$.
We prefer to keep (28) generic and the $i$ factors in the gauge
interactions for Majorana neutrinos with $\eta _{CP} = - i$.

If $U_L$ is of the form in Eq. (21), the gauge interactions in
Eqs. (24,25) conserve CP. Any extra phase gives rise to CP violation.
Hence, the number of CP violating phases is equal to the number of
independent phases in $n_L$ rows of a $(n_L + n_R)\times (n_L + n_R)$
unitary matrix ($n_L (n_L + n_R) - \frac{n_L (n_L - 1)}{2}$) minus
the number of phases which can be absorbed in the charged lepton mass
eigenstate definition ($n_L$) [5]:
\begin{equation}
n_L (n_L + n_R) - \frac{n_L (n_L - 1)}{2} - n_L =
\frac{n_L(n_L+2n_R-1)}{2}.
\end{equation}
{\bf 2.4. CP invariant constraints on the lepton mass matrices}\\

In this Section we derive from Eqs. (12) neccesary conditions for
CP conservation which are independent of the weak basis and do not
require to know the unitary matrices involved in the definition
of the CP transformation. If these conditions, which are simple
functions of the mass matrices, are not satisfied, CP is violated.
That they are not only necessary but sufficient has to be proven
case by case. It looks feasible (and is proven) in (the)
simple(st) cases only. This formulation provides (and we obtain) the
factors suppressing CP violating observables. Their knowledge will
allow for a discussion of the size of CP violating effects.

Motivated by Eqs. (12) and the quark case we classify the products of
the mass matrices $M_{l,L,D,R}$, as well as their sums,
in three classes $G_L, G_R^l, G_R^{\nu}$, depending under
which unitary matrix $V_L, V_R^l, V_R^{\nu}$ they transform, respectively:
\begin{equation}
\begin{array}{c}
V^{\dagger}_L G_L V_L = G_L^* , \\
V^{l \dagger}_R G^l_R V_R^l = G_R^{l *} , \\
V^{\nu \dagger}_R G^{\nu}_R V^{\nu}_R = G_R^{\nu *} .
\end{array}
\end{equation}
To these classes belong:
\begin{equation}
\begin{array}{ccc}
\{ G_L \} & = & \{A_{L1} = M_lM_l^{\dagger};
A_{L2} = M_LM_L^{\dagger}; A_{L3} = M_DM_D^{\dagger}; \\
 & & A_{Li}A_{Lj}, i,j = 1,2,3;
M_LM_l^*M_l^TM_L^{\dagger}; \\
 & & M_LM_D^*M_D^TM_L^{\dagger}; M_LM_D^*M_RM_D^{\dagger};
M_DM_R^{\dagger}M_RM_D^{\dagger}; \\
 & & {\rm and \ higher \ order \ products; \ and \ sums } \} , \\
\{ G_R^l \} & = & \{A_l = M_l^{\dagger}M_l; A_l^2;
M_l^{\dagger}M_LM_L^{\dagger}M_l;
M_l^{\dagger}M_DM_D^{\dagger}M_l; \\
 & & {\rm and \ higher \ order \ products; \ and \ sums } \} , \\
\{ G_R^{\nu} \} & = & \{A_{\nu 1} = M_D^{\dagger}M_D;
A_{\nu 2} = M_R^{\dagger}M_R; A_{\nu i}A_{\nu j}, i,j = 1,2; \\
 & & M_D^{\dagger}M_LM_L^{\dagger}M_D;
M_D^{\dagger}M_LM_D^*M_R; M_R^{\dagger}M_D^TM_D^*M_R; \\
 & & {\rm and \ higher \ order \ products; \ and \ sums } \} .
\end{array}
\end{equation}
Now observing that the trace and the determinant of any element
of these classes is invariant under unitary transformations, and
then under weak basis transformations, and that Eqs. (30), which
follow from CP conservation, implies that these traces and determinants
are real, we can write a set of necessary conditions for CP conservation
which are weak basis invariant:
\begin{equation}
\begin{array}{c}
Im Tr (G) = 0 , \\
Im Det (G) = 0 ,
\end{array}
\end{equation}
where $G$ is any element of $\{ G_L \}, \{ G_R^l \} \ {\rm or } \
\{ G_R^{\nu} \}$ in Eqs. (31).

Eqs. (32) are the corner stone of our analysis. They apply to any
number of standard families $n_L$ and right-handed neutrinos $n_R$.
Although these conditions are all necessary, they are not all
independent. To find a set of conditions which are also sufficient
seems to be difficult in general. We shall obtain such a subset
of necessary and sufficient CP conserving conditions for some simple
(lowest $n_L$ and $n_R$) cases only.

We will find a set of necessary and sufficient conditions for
$n_L + n_R < 3$. In each model we state
the number of CP violating phases, Eq. (29); the set of necessary
(and sufficient) CP constraints; a parametrization of $M_{\nu}$ in the
convenient weak basis where $M_l$ and $M_R$ are diagonal with
real (and positive $M_l$) elements; the expressions of the CP constraints
for this parametrization; the proof that if these invariants vanish CP is
conserved ($M_l$ and $M_{\nu}$ can be made real); and the expressions
of these conditions as functions of physical observables
(masses and mixing angles). The cases with $n_L + n_R = 3$ are also
discussed. \\

$\bullet \ n_L = 0, n_R; \ n_L = 1, n_R = 0$: \\

There is no CP violation (and, of course, no CP violating phase)
in these models.\\

$\bullet \ n_L = 1, n_R = 1$: \\

In this case there is one CP violating parameter. A necessary and
sufficient condition for CP conservation is
\begin{equation}
\Delta_{11} = Im Tr ( M_D^{\dagger} M_L M_D^* M_R ) = 0.
\end{equation}
$M_{D,L,R}$ are one-dimensional in this model and we can write
\begin{equation}
M_{\nu} = {m_L \ \  ae^{i\alpha} \choose ae^{i\alpha} \ \ m_R} ,
\end{equation}
in the convenient weak basis, which is completely specified requiring
$a \ge 0,\ \alpha \in [0, \frac {\pi}{2} )$. In this parametrization
\begin{equation}
\Delta_{11} = - m_L m_R a^2 sin (2\alpha ) .
\end{equation}
The vanishing of $\Delta_{11}$ is apparently a sufficient condition for
CP conservation: $M_{\nu}$ is real for $a$ or $\alpha = 0$, and it can
be made real by a field redefinition for $m_L$ or $m_R = 0$, choosing
$W_L = e^{-i\alpha}$ or $W_R^{\nu} = e^{i\alpha}$, respectively, in
Eq. (14). As a function of physical observables
\begin{equation}
\Delta_{11} = m_1 m_2 (m_2^2 - m_1^2) Im ( U_{11}^{*\ 2} U_{12}^2 ) .
\end{equation}
Thus, CP is conserved if there is a massless neutrino ($m_1$ or
$m_2 = 0$), or the neutrinos are degenerate ($m_1 = m_2$). On the
other hand, any CP violating effect is proportional to
$Im ( U_{11}^{*\ 2} U_{12}^2 )$, where $U_{11} (U_{12})$ is
the element in the mixing matrix in Eq. (24) fixing the charged
coupling of the charged lepton to the neutrino of mass $m_1 (m_2)$. \\

$\bullet \ n_L = 2, n_R = 0$: \\

This model has deserved some attention [9]. It has one CP violating
parameter and one necessary and sufficient condition for CP
conservation is
\begin{equation}
\Delta_{20} = Im Det ( M_L M_l^* M_l^T M_L^{\dagger} -
M_L M_L^{\dagger} M_l M_l^{\dagger} ) = 0.
\end{equation}
In the basis where $M_l$ is diagonal with charged lepton masses
$m_e, m_{\mu}$ and
\begin{equation}
M_{\nu} = M_L = {a \ \  be^{i\beta} \choose be^{i\beta} \ \ c} ,
\ a, b \ge 0,\ \beta \in [0, {\frac {\pi}{2}}),
\end{equation}
\begin{equation}
\Delta_{20} = - ( m_{\mu}^2 - m_e^2)^2 acb^2 sin ( 2 \beta ).
\end{equation}
It is easy to prove that $\Delta_{20} = 0$ is a sufficient condition
for CP conservation: $M_{\nu}$ is real
for $b$ or $\beta = 0$, and it can be made real for $a = 0$; and
$c = 0$, choosing in Eqs. (13-15)
\begin{equation}
W_L = W_R^l = {e^{-i\beta} \  0 \choose \ 0 \ \ \ 1};\ {\rm and}\
W_L = W_R^l = {1 \ \ \ 0 \ \choose 0 \ e^{-i\beta}},
\end{equation}
respectively. $\Delta_{20}$ also vanishes for $m_e = m_{\mu}$, but
in this case $M_l$ can be diagonal and $W_L = W_R^l$ be still an arbitrary
unitary matrix. Hence, $M_{\nu}$ can be made not only real but diagonal
and positive by an appropriate choice of this unitary transformation:
\begin{equation}
W_L M_{\nu} W_L^T = {m_1 \ \ \ \ \ \choose \ \ \ \ \ m_2},
\end{equation}
where $m_{1,2}$ are the neutrino masses. As a function of physical
observables
\begin{equation}
\Delta_{20} = m_1m_2({m^{2}_{2}} - {m^{2}_{1}})
{({m^{2}_{\mu}} - {m^{2}_{e}})}^2
Im ( U^{*\ 2}_{11}{U^{2}_{12}} ).
\end{equation}
Hence, CP violation requires two massive ($m_{1,2} \ne 0$)
and non-degenerate ($m_1 \ne m_2$) neutrinos, as well as
two non-degenerate charged leptons ($m_e \ne m_{\mu}$).
As for $n_L = n_R = 1$
any CP violating effect is proportional to
$Im ( U_{11}^{*\ 2} U_{12}^2 )$, where $U_{11} (U_{12})$ is
the element in the mixing matrix in Eq. (24) fixing the charged
coupling of one charged lepton, let say $e$, to the neutrino
of mass $m_1 (m_2)$. (Notice that $Im ( U_{11}^{*\ 2} U_{12}^2 )
= Im ( U_{21}^{*\ 2} U_{22}^2 )$.) \\

$\bullet \ n_L + n_R \ge 3$: \\

There are two CP violating phases for $n_L = 1, n_R = 2$. Two necessary
and sufficient conditions for CP conservation are
\begin{equation}
\begin{array}{c}
\Delta_{12}^{(1)} = Im Tr ( M_D^{\dagger} M_L M_D^* M_R ) = 0, \\
\Delta_{12}^{(2)} = Im Det ( M_R^{\dagger} M_D^T M_D^* M_R -
M_D^{\dagger} M_D M_R^{\dagger} M_R ) = 0.
\end{array}
\end{equation}

There are three CP violating phases for $n_L = 2, n_R = 1$.
It can be proved, after some work, that
\begin{equation}
\begin{array}{c}
\Delta_{21}^{(1)} = Im Det (M_L M_l^* M_l^T M_L^{\dagger} -
M_L M_L^{\dagger} M_l M_l^{\dagger}) = 0, \\
\Delta_{21}^{(2)} = Im Det (M_D M_D^{\dagger} M_L M_L^{\dagger} -
M_L M_D^* M_D^T M_L^{\dagger}) = 0, \\
\Delta_{21}^{(3)} = Im Det (M_D M_D^{\dagger} M_l M_l^{\dagger} -
M_L M_D^* M_D^T M_L^{\dagger}) = 0, \\
\Delta_{21}^{(4)} = Im Det (M_D M_D^{\dagger} M_l M_l^{\dagger} -
M_l M_l^{\dagger} M_L M_L^{\dagger}) = 0, \\
\Delta_{21}^{(5)} = Im Det (M_L M_l^* M_l^T M_L^{\dagger} -
M_D M_D^{\dagger} M_l M_l^{\dagger}) = 0, \\
\Delta_{21}^{(6)} = Im Tr (M_D^{\dagger} M_L M_D^* M_R) = 0, \\
\Delta_{21}^{(7)} = Im Tr (M_l M_l^{\dagger} M_D M_R^* M_D^T M_l^* M_l^T
M_L^{\dagger}) = 0, \\
\Delta_{21}^{(8)} = Im Tr ((M_l M_l^{\dagger})^2 M_D M_R^* M_D^T
(M_l^* M_l^T)^2 M_L^{\dagger}) = 0, \\
\end{array}
\end{equation}
form a set of necessary and sufficient conditions for CP conservation.

For $n_L = 3, n_R = 0$, although there are three CP breaking phases
as for $n_L = 2, n_R = 1$, it seems difficult to find a subset
of sufficient conditions for CP conservation. Analogously to the quark
case [3] the obstacle, which is generic for large(r) $n_L, n_R$,
is the non-linearity of conditions (32) on the CP breaking phases.
We expect to handle this case, as well as the models with $n_L =
n_R = 2, 3$, with a computer. \\

{\bf 3. Extended models with left-right symmetry}\\

The analysis of left-right models parallels the analysis of
extended models with extra right-handed neutrinos only. The
number of left-handed and right-handed neutrinos is now the
same $n_L = n_R = n$. There are also extra charged and neutral
currents which further constrain the CP transformations. \\

{\bf 3.1. CP invariance}\\

The gauge interactions can be written in this case [10]:
\begin{equation}
{\cal L}_{CC} = - \frac{g}{\sqrt{2}} ( \bar {\nu}_L
\gamma^{\mu} l_L W^+_{L \mu} + \kappa \bar {\nu}_R
\gamma^{\mu} l_R W^+_{R \mu} ) + h.c.,
\end{equation}
and
\begin{equation}
\begin{array}{c}
{\cal L}_{NC} = e \ ( \bar l_L \gamma^{\mu} l_L +
\bar l_R \gamma^{\mu} l_R ) \ A_{\mu} \\
+ \ \frac{g}{2 cos \theta _W} \ \left( - \bar {\nu}_L \gamma^{\mu} \nu _L
+ ( 1 - 2 sin^2 \theta _W ) \ \bar l_L \gamma^{\mu} l_L
- 2 sin^2 \theta _W \ \bar l_R \gamma^{\mu} l_R \right) \ Z_{\mu} \\
- \ \frac{e}{2 cos \theta _W} \frac{1}{\alpha} \ ( \bar {\nu}_L
\gamma^{\mu} \nu _L + (1 + \alpha ^2) \bar {\nu}_R \gamma^{\mu} \nu _R
+ \bar l_L \gamma^{\mu} l_L + (1 - \alpha ^2) \bar l_R \gamma^{\mu}
l_R ) \ Z_{LR\ \mu},
\end{array}
\end{equation}
with $\alpha = \sqrt {\kappa ^2 cot ^2 \theta _W - 1}$ real. $W_{L,R}^{\pm}$
are the charged gauge bosons associated to $SU(2)_{L,R}$; and $Z, Z_{L,R}$
are the two heavy neutral gauge bosons associated to the standard model and
its left-right $(LR)$ extension, respectively. Whereas $l_{L,R}$ and
${\nu}_{L,R}$ are $n_L = n_R$ component vectors in flavour space describing
left-, right-handed charged leptons and left-, right-handed neutrinos.
${\cal L}_{CC,NC}$ are now left invariant by any CP transformation
\begin{equation}
\begin{array}{c}
\nu_{L} \longrightarrow V_{L}C{\nu^{*}_{L}}, \\
l_L \longrightarrow V_{L}C{l^{*}_{L}}, \\
l_{R} \longrightarrow {V_{R}}C{l^{*}_{R}}, \\
\nu_{R} \longrightarrow V_{R}C{\nu^{*}_{R}},
\end{array}
\end{equation}
where $V_{L,R}$ are $n \times n$ unitary matrices. On the other hand,
the mass matrices $M_{l,L,D,R}$ are all $n \times n$. The invariance
of the mass terms in Eq. (10) under the CP transformations in Eq. (47)
implies
\begin{equation}
\begin{array}{c}
{V^{\dagger}_{L}}M_{l}{V_{R}} = {M^{*}_{l}}, \\
{V^{\dagger}_{L}}M_{L}{V^{*}_{L}} = {M^{*}_{L}}, \\
{V^{\dagger}_{L}}M_{D}{V_{R}} = {M^{*}_{D}}, \\
{V^{T}_{R}}M_{R} V_R = {M^{*}_{R}}.
\end{array}
\end{equation}
These conditions are necessary and sufficient for CP conservation in
left-right symmetric models. \\

{\bf 3.2. Weak basis independence}\\

As before conditions (48) are weak basis independent: Eqs. (13-15)
apply but with $W_R^l = W_R^{\nu} = W_R, \ V_R^l = V_R^{\nu} = V_R$
and ${V'}_R^{\ l} = {V'}_R^{\ \nu} = {V'}_R$. \\

{\bf 3.3. CP conserving gauge interactions in the mass eigenstate basis
and counting of CP breaking phases in the lepton sector}\\

The freedom to choose the weak basis allows to assume that $M_l$
is diagonal. In this basis Eq. (48) for $M_l$ (for non-degenerate
charged lepton masses) implies
\begin{equation}
(V_L)_{ij} = (V_R)_{ij} = e^{{i{\delta_i}}}\delta_{ij}.
\end{equation}
Then Eqs. (48) for $M_{L,D,R}$ are satisfied and CP conserved if and
only if
\begin{equation}
\begin{array}{c}
(M_L)_{ij} = \  (M^r_L)_{ij} \  e^{ i \frac{1}{2} ( \delta_i +
\delta_j)  } , \\
(M_D)_{ij} = \  (M^r_D)_{ij} \  e^{i\frac{1}{2}(\delta_i - \delta_j)}, \\
(M_R)_{ij} = \  (M^r_R)_{ij} \  e^{ - i \frac{1}{2} ( \delta_i +
\delta_j)  } ,
\end{array}
\end{equation}
with $(M^r_{L,D,R})_{ij}$ real.
The number of possible CP violating phases is larger in left-right
models because the definition of a CP transformation is less
general. To write the CP conserving gauge interactions for left-right models
in the mass eigenstate basis, we follow the same steps as in Section 2.3.
Now ($\Pi = \Sigma ^* $) [11]
\begin{equation}
U  =  {\Sigma ^* \ \ 0 \choose \ 0 \ \ \Sigma }
O \Lambda ,
\end{equation}
and
\begin{equation}
\begin{array}{c}
{\cal L}_{CC} =
- \frac{g}{\sqrt{2}} ( \bar N_L U^{\dagger}_L
\gamma^{\mu} l_L W^+_{L\ \mu} + \kappa \bar N_R U^{\dagger}_R
\gamma^{\mu} l_R W^+_{R\ \mu} ) + h.c. \\
= - \frac{g}{\sqrt{2}} ( \bar N_L \Lambda O^T_L
\gamma^{\mu} \Sigma ^* l_L W^+_{L\ \mu} + \kappa \bar N_R \Lambda ^*
O^T_R \gamma^{\mu} \Sigma ^* l_R W^+_{R\ \mu} ) + h.c. ,
\end{array}
\end{equation}
where $N_{L,R} = U^T \psi _{L,R}$ are the neutrino mass eigenstates.
As in Eq. (24) the $\Sigma ^*$ phases can be absorbed in the charged
lepton mass eigenstates defining then equal to $e^{-i\frac {\delta _i}
{2}} l_{L,R\ i}$. The neutral current lagrangian in Eq. (46) remains
unchanged for the charged lepton mass eigenstates; whereas for the neutrinos
can be written
\begin{equation}
\begin{array}{c}
{\cal L}_{NC} =
- \frac{g}{2 cos \theta _W} \bar N_L U^{\dagger}_L
\gamma^{\mu} U_L N_L Z_{\mu} \\
- \frac{e}{2 cos \theta _W} \frac{1}{\alpha}
( \bar N_L U^{\dagger}_L \gamma^{\mu} U_L N_L + (1 + \alpha ^2)
\bar N_R U^{\dagger}_R \gamma^{\mu} U_R N_R ) Z_{LR\ \mu} \\
= - \frac{g}{2 cos \theta _W} \bar N_L \Lambda O^T_L O_L
\gamma^{\mu} \Lambda ^* N_L Z_{\mu} \\
- \frac{e}{2 cos \theta _W} \frac{1}{\alpha}
( \bar N_L \Lambda O^T_L O_L \gamma^{\mu} \Lambda ^* N_L
+ (1 + \alpha ^2)
\bar N_R \Lambda ^* O^T_R O_R \gamma^{\mu} \Lambda N_R ) Z_{LR\ \mu} .
\end{array}
\end{equation}

The number of possible CP violating phases in ${\cal L}_{CC,NC}$
(Eqs. (52,53)) for a general $2n \times 2n$ unitary matrix
$U = {U^*_L \choose U_R}$ is equal to the number of independent
phases in $U \ (\frac{2n(2n+1)}{2})$ minus the number of phases
which can be absorbed in the charged lepton mass eigenstate
definition ($n$) [5]:
\begin{equation}
n (2n + 1) - n = 2n^2.
\end{equation}
{\bf 3.4. CP invariant constraints on the lepton mass matrices }\\

As in Section 2.4. we shall derive necessary conditions for CP
conservation which are independent of the weak basis and do not
require to know the unitary matrices involved in the definition
of the CP transformation. That a subset of them is also sufficient
has to be proven case by case.

The products of the mass matrices $M_{l,L,D,R}$, as well as their
sums, can be classified in two classes $G_{L,R}$, which under
$V_{L,R}$ transform, respectively:
\begin{equation}
\begin{array}{c}
V^{\dagger}_L G_L V_L = G_L^* , \\
V^{\dagger}_R G_R V_R = G_R^* .
\end{array}
\end{equation}
To these classes belong:
\begin{equation}
\begin{array}{ccc}
\{ G_L \} & = & \{A_{L1} = M_lM_l^{\dagger};
A_{L2} = M_LM_L^{\dagger}; A_{L3} = M_DM_D^{\dagger}; \\
 & & A_{L4} = M_lM_D^{\dagger};
A_{L5} = M_DM_l^{\dagger}; A_{Li}A_{Lj}, \\
 & & i,j = 1,2,...,5;
B_i B_j^{\dagger} , i,j = 1,2,...,4;  \\
 & & {\rm and \ higher \ order \ products; \ and \ sums } \} , \\
\{ G_R \} & = & \{A_{R1} = M_l^{\dagger}M_l;
A_{R2} = M_D^{\dagger}M_D; A_{R3} = M_R^{\dagger}M_R; \\
 & & A_{R4} = M_l^{\dagger}M_D;
A_{R5} = M_D^{\dagger}M_l; A_{Ri}A_{Rj}, \\
 & & i,j = 1,2,...,5;
B_i^T B_j^* , i,j = 1,2,...,4; \\
 & & {\rm and \ higher \ order \ products; \ and \ sums } \} ,
\end{array}
\end{equation}
where $B_1 = M_lM_R^{\dagger};
\ B_2 = M_LM_l^*; \ B_3 = M_LM_D^*; \ B_4 = M_DM_R^{\dagger}$.
As before, Eqs. (32) but with $G$ any element of $\{ G_L \}$
or $\{ G_R \}$ in Eq. (56), are a set of necessary conditions for CP
conservation which are weak basis invariant. They follow from Eqs. (55)
and the invariance of the trace and the determinant under unitary
transformations. We shall obtain a subset of sufficient conditions
for $n = 1$. For $n = 2$ there are eight CP violating phases. A
set of necessary and sufficient conditions will include the
constraints in Eqs. (43,44) and many more,
requiring its analysis a long casuistry. Besides, for larger $n$
there is the same difficulty to
find a subset of sufficient constraints as for extended models
with only extra right-handed neutrinos:
conditions (32) are non-linear on the CP breaking phases.
We expect to handle these cases with the help of a computer. \\

$\bullet \ n = 1$: \\

There are two CP violating parameters in this case (which is the
simplest one). Two necessary and sufficient conditions for CP
conservation are
\begin{equation}
\begin{array}{c}
\Delta_1^{(1)} = Im Tr ( M_l M_D^{\dagger} ) = 0, \\
\Delta_1^{(2)} = Im Tr ( M_l M_R^{\dagger} M_l^T M_L^{\dagger}
- M_L M_D^* M_R M_D^{\dagger} ) = 0.
\end{array}
\end{equation}
The neutrino mass matrix can be written in the convenient weak
basis where $M_{l,L}$, which are one-dimensional, are real
and $M_l$ is also positive
\begin{equation}
M_{\nu} = {m_L \ \  ae^{i\alpha} \choose ae^{i\alpha} \ m_Re^{i\phi}} ,
\ a \ge 0,\ \alpha , \phi \in [0, \pi ).
\end{equation}
In this parametrization
\begin{equation}
\begin{array}{c}
\Delta_1^{(1)} = - m_e a sin ( \alpha ) , \\
\Delta_1^{(2)} = - m_L m_R ( m_e^2 sin ( \phi ) + a^2
sin ( \phi - 2\alpha ) ) ,
\end{array}
\end{equation}
with $m_e$ the charged lepton mass. The vanishing of $\Delta_1^{(1,2)}$
is sufficient for CP conservation: $M_{\nu}$ is real for
$a, m_R = 0$; $\alpha , m_R = 0$; and $\alpha , \phi = 0$, and it can
be made real by a field redefinition for $m_e, a = 0$ or
$m_e = 0, \phi - 2\alpha = 0, -\pi$; $m_L, a = 0$ or $m_L, \alpha = 0$;
$m_e, m_L = 0$; and $m_e, m_R = 0$, choosing $W_R = e^{i\frac {\phi}{2}}$;
$W_L = W_R = e^{i\frac {\phi}{2}}$; $W_L = e^{i(\frac {\phi}{2}-\alpha )},\
W_R = e^{i\frac {\phi}{2}}$; and $W_R = e^{i\alpha }$, respectively.
As a function of physical observables
\begin{equation}
\begin{array}{c}
\Delta ^{(1)}_1 = m_e Im ( m_1 U_{11} U_{21} + m_2 U_{12} U_{22} ) , \\
\Delta ^{(2)}_1 = m_e^2 Im (( m_1 U_{11}^2 + m_2 U_{12}^2 )
( m_1 U_{21}^2 + m_2 U_{22}^2 )) \\
- m_1m_2({m^{2}_{2}} - {m^{2}_{1}}) Im ( U^{*\ 2}_{11}{U^{2}_{12}} ) ,
\end{array}
\end{equation}
where $m_{1,2}$ are the neutrino masses and $U_{ij}$ are the $U =
{U^*_L \choose U_R}$ matrix elements fixing the charged coupling
of the left- (right-) handed charged lepton $i = 1 (2)$ to the
neutrino of mass $m_j$ (see Eq. (52)). This model has two
(independent) physical phases, which we associate to $P =
m_1 U_{11} U_{21} + m_2 U_{12} U_{22}$ and $Q = ( m_1 U_{11}^2 +
m_2 U_{12}^2 ) ( m_1 U_{21}^2 + m_2 U_{22}^2 )$ in Eq. (60),
respectively. Then, any CP violating effect can be written as
a function of $P$ and/or $Q$. In particular, the second term of
$\Delta ^{(2)}_1$ in (60) is equal to $- Im (P^2 Q^*)$. \\

{\bf 4. CP violating effects and conclusions}\\

In this paper we study the formulation of CP violation in the lepton
sector in extended models with right-handed neutrinos, without and with
left-right symmetry. We obtain necessary conditions for CP conservation
for any number of standard families $n_L$ and right-handed neutrinos
$n_R$ (in left-right symmetric models $n_L = n_R = n$). These are
independent of the weak basis used to express the charged lepton and
the neutrino mass matrices. However, they are not all
independent, and in general it appears difficult to find a subset
of sufficient conditions . Proceeding case by case we do find such a
set of necessary and sufficient conditions for the simplest cases
(lowest $n_L, n_R$): $n_L = 0, n_R$; $n_L = 1, n_R = 0;\  n_L = 1, n_R = 1;
\  n_L = 2, n_R = 0; \ n_L = 1, n_R = 2;\  n_L = 2, n_R = 1$,
and $n = 1$. The proof
takes profit of the freedom to choose the weak basis. This is fixed
requiring that the charged lepton mass matrix be diagonal with real
positive eigenvalues and the
neutrino mass matrix have as many diagonal blocks and real (positive)
diagonal elements as possible. For cases with larger $n_L, n_R$ we
have to rely on a computer.

This invariant formulation allows for deciding more easily if CP is
conserved because the CP conserving constraints can be simply
calculated in any weak basis. It may also help to
understand the origin of CP violation if a pattern of leptonic
mass matrices is distinguished for some physical reason. It
can be used as a guide for model buildig. However, the
main practical application of constructing such a set of necessary and
sufficient conditions for CP conservation follows from observing
that any CP violating effect is proportional to weak basis invariants,
and then to the invariant factors appearing in these constraints.
This permits to discuss the possible suppression factors and then the
size of the CP violating observables.

As an example we can compare the cases $n_L = n_R = 1$, without, and $n = 1$,
with left-right symmetry. CP is violated for $n_L = n_R = 1$ if and only if
(see Section 2.4)
$$
Im (( m_1 U_{11} U_{21} + m_2 U_{12} U_{22} )^2
(( m_1 U_{11}^2 + m_2 U_{12}^2 ) ( m_1 U_{21}^2 + m_2 U_{22}^2 ))^*)
\neq 0  ,
$$
and for $n = 1$ if and only if (see Section 3.4)
$$ m_e Im ( m_1 U_{11} U_{21} + m_2 U_{12} U_{22} ) \neq 0 ,
\ {\rm and/or } $$
$$ m_e^2 Im (( m_1 U_{11}^2 + m_2 U_{12}^2 )
( m_1 U_{21}^2 + m_2 U_{22}^2 ))
- m_1m_2({m^{2}_{2}} - {m^{2}_{1}})
Im ( U^{*\ 2}_{11}{U^{2}_{12}} ) \neq 0 .
$$
Therefore, CP violation can be larger in left-right models because
any CP violating effect in the presence of right-handed currents is
a function of
$P = m_1 U_{11} U_{21} + m_2 U_{12} U_{22}$ and/or
$Q = ( m_1 U_{11}^2 + m_2 U_{12}^2 ) ( m_1 U_{21}^2 + m_2 U_{22}^2 )$;
whereas a CP violating observable in their absence is a function
of the product $P^2 Q^*$ only, whose imaginary part above is equal to
$m_1m_2({m^{2}_{2}} - {m^{2}_{1}}) Im ( U^{*\ 2}_{11}{U^{2}_{12}} )$.
This is generic.
In left-right models CP can be violated in left-handed as well as in
right-handed currents.
That CP violation can be larger
in left-right models can be seen when producing two heavy neutrinos
$N_1N_2$ at $e^+e^-$ [11]. \\

{\bf Acknowledgements} \\

This work was partially supported by CICYT under contract AEN94-0936.
F. A. was also supported by the Junta de Andaluc\'\i a and by
the European Union under contract CHRX-CT92-0004 and M. Z. by
the Curie Sklodowska grant MEN/NSF 93-145.

\end{document}